\documentstyle[12pt]{article}

\begin{document}

\title{Lepton Flavor Violation\\ and Fermion Masses\footnote{Talk presented at
the Workshop on Physics at the First Muon Collider and at the Front End of a Muon
Collider, November 6-9,1997, FNAL.}}

\author{{\bf Stuart Raby}\\ The Ohio State University\\174 W. 18th Ave \\
Columbus OH 43210 \\ raby@mps.ohio-state.edu}
\date{OHSTPY-HEP-T-97-025}
\maketitle

\begin{abstract}
Lepton flavor violation may be a signature of ``GUT scale" physics, if the 
messenger scale for SUSY breaking is above the ``GUT scale."   We elaborate 
on the details of this
simple statement in the following talk.
\end{abstract}

\section*{Introduction}
The minimal supersymmetric standard model [MSSM] is defined by its spectrum
and interactions, i.e. the minimal particle spectrum necessary for a 
self-consistent extension of the standard  model, along with R parity so 
that the only interactions are those in the standard model or supersymmetric
extensions thereof. Even with these constraints the theory in principle has 
many unknown parameters.   These are associated  with soft SUSY breaking 
parameters defined at a messenger scale, $M$.  In minimal 
supergravity\cite{sugra} the messenger scale 
$M = M_{Pl} \sim 10^{18}$ GeV.  In this case SUSY breaking occurs in a hidden 
sector and is transmitted to the visible sector via  gravitational interactions.
It results in 5 soft SUSY breaking parameters, a {\em universal scalar mass}
$m_0$,  a universal gaugino mass $M_{1/2}$, a supersymmetric Higgs mass 
parameter $\mu$ (which in some theories is only generated once SUSY is broken), 
the Higgs scalar  mass $B \mu$ and a {\em universal soft trilinear interaction
parameter} $A$.  In gauge-mediated SUSY breaking\cite{lsgauge,hsgauge,dn}, on 
the otherhand, the messenger scale is typically much less than $M_{Pl}$.  In
this case, scalars with common gauge charges are degenerate and $A$ vanishes
at tree level.  In 
this talk we consider minimal supergravity SUSY breaking, unless
otherwise stated. Finally in the MSSM, as in the standard
model, {\em neutrinos are massless}.

The MSSM as defined above is a symmetry limit.   Individual lepton numbers,
$L_e, \; L_{\mu},\; L_{\tau}$, are conserved.  Thus processes such as
$\mu \rightarrow e \gamma$,  $\mu \rightarrow 3 \, e$, $ \mu \rightarrow
e$ conversion or $\tau \rightarrow \mu \gamma$ are forbidden.
The experimental branching ratios for these processes are bounded by\cite{pdg}
$B(\mu \rightarrow e \gamma) \leq 5 \times 10^{-11}$, $B(\mu \rightarrow 3 \,
e) \leq 1 \times 10^{-12}$,  $B(\mu\; _{22}^{48}T_i \rightarrow e \;
_{22}^{48}T_i) \leq 4.3 \times 10^{-12}$ and $B(\tau \rightarrow \mu \gamma)
\leq 4.2 \times 10^{-6}$.

These strong constraints have two significant consequences.

\begin{enumerate}
\item {\bf Possible non-universal scalar masses or soft trilinear parameters are
severely constrained\cite{dg,ggms}.}   For example, define
\begin{eqnarray}
\delta_{ij}^{\bar e} & \equiv {\Delta_{ij}^{\bar e} \over \tilde m^2} &  \nonumber
\\
\delta_{ij}^{LR} & \equiv {\Delta_{ij}^{LR} \over \tilde m^2} &   
\end{eqnarray}
where  $\Delta_{ij}^{\bar e} \;(\Delta_{ij}^{LR})$ is the off-diagonal 
mass squared term for right-handed (left-to-right handed) scalar leptons in a
superbasis where lepton masses are diagonal ($i, j$ are flavor indices).
 Then typical constraints\cite{ggms}  are, 
\begin{eqnarray}
\delta_{12}^{\bar e} & < 4.3 \times 10^{-3} \; \left({\tilde m_{\bar e}(GeV)
\over 100}\right)^2 &   \label{eq:bound}
\end{eqnarray}
(from $\mu \rightarrow e \gamma$ with  $({m_{\tilde \gamma} \over \tilde
m_{\bar e}})^2 = 0.3$), or
\begin{eqnarray}
\delta_{ij}^{LR} & < 1.5 \times 10^{-6} \; \left({\tilde m(GeV)
\over 100}\right)^2 &   
\end{eqnarray}

\item {\bf Lepton flavor violation[LFV] is sensitive to ``GUT scale"
physics\cite{hkr,bm}.}   Although ``GUT scale" physics could easily violate 
flavor symmetries, one might suspect these flavor violations to  
be suppressed by powers of $1/M_G$. 
This is not the case however.  As shown by Hall et al.
\cite{hkr}  flavor violation in the lepton sector can be induced at the GUT 
scale due to RG running from $M$ to $M_G$.  Moreover, this flavor violation 
enters as a boundary condition in the slepton mass matrices; hence it is not 
suppressed by inverse powers of $M_G$.  

As an illustration of this phenomenon, consider
a generic GUT-like theory with heavy states $X,\; Y,\; Z$ with mass
$M_I \sim M_G$ and the standard model states $F_i = \{ Q_i,\; \bar u_i,\; \bar d_i,\;
L_i,\; \bar e_i \} $.   Assume some new interactions between the scales $M_I$
and $M$ given by
\begin{eqnarray}
 &  \lambda_{ij} \; F_i \; F_j \; X  & + \;\; k_i \; F_i \; Y \; Z 
\end{eqnarray}

As a consequence of renormalization group running, we find at $M_I$.\footnote{Universal 
scalar masses at $M$ are assumed in all analyses.}
\begin{eqnarray}
\delta^F_{ij} & \sim \; - ( (\lambda^{\dagger} \, \lambda)_{ij} + k_i^{\dagger} \;
k_j ) \; ln {M \over M_I} &
\end{eqnarray} 

Of course, the numerical value depends on the scale $M_I$ and 
the magnitude of the Yukawa couplings $\lambda_{ij},\; k_i$.  In the rest of 
this talk we consider three different possible contributions to lepton flavor 
violating interactions emanating from ``GUT scale" physics.  In all three cases, $M_I$ is
the scale where the structure of the fermion mass hierarchy is generated.

\begin{itemize}
\item[A.] \hspace{.5in}{\bf Adding neutrino masses to the MSSM}
\item[B.] \hspace{.5in}{\bf GUTs and the Third family yukawa couplings}
\item[C.] \hspace{.5in}{\bf Family mass hierarchy and the FN mechanism}
\end{itemize}
\end{enumerate}

\subsection*{A. --- Adding neutrino masses to the MSSM}

Consider adding to the MSSM some right-handed neutrinos; one for each family.
The most general renormalizable superspace potential including the right-handed
neutrinos is given by
\begin{eqnarray}
W & = \;  \lambda_{ij}^{\nu} \; \bar \nu_i \; L_j  \; h  \;\; + \;\; M_{ij} \;
\bar \nu_i \; \bar \nu_j &  
\end{eqnarray}
where  $M_{ij} = \delta_{ij} M_i$ ($\delta_{ij}$ is a Kronecker delta) and we 
work in a basis where charged lepton yukawa couplings are diagonal.
In this case, the scale $M_I$ is given by $M_I = \min\{M_i\} \gg M_Z$.  As 
long as $\det M \neq 0$, this theory results in 3 light majorana neutrinos 
(predominantly left-handed) and 3 superheavy majorana neutrinos (predominantly 
right-handed). RG running leads to radiative mass corrections of the form 
\cite{bm,hmtty}
\begin{eqnarray}
\delta_{ij}^L  & \sim \; - (\lambda^{\nu \; \dagger} \; \lambda^{\nu})_{ij} \; \ln({M \over
M_I}) &
\end{eqnarray}

A recent analysis by Hisano et al.\cite{hmtty}
 takes $\lambda_{ij}^{\nu} = \lambda_i^u \; V^{CKM}_{ij}  $ where $\lambda_i^u $ 
are the diagonal up quark yukawa couplings and $V^{CKM}$ is the CKM matrix.  This
form for $\lambda_{ij}^{\nu}$ is suggested by SO(10) GUTs.  A value for
$M_I$ of order $10^{12}$ GeV was also assumed.  With this value, the tau neutrino has
mass of a few eV and thus it makes a good hot dark matter
candidate in a universe with hot + cold dark matter.  Branching ratios  for LFV processes
are obtained which are below the experimental bounds but close enough to be 
observable in future LFV experiments at Los Alamos or PSI.

\subsection*{B. --- GUTs and the Third family yukawa couplings}

Quark flavor is not conserved; this is the essense of CKM mixing.  Since
GUTs relate quarks and leptons, it is not surprising that GUT interactions also violate
lepton flavor. 

For example, consider  a simple SUSY SU(5) model with quarks and leptons in
the  $10_i \supset \{ Q_i, \; \bar u_i, \; \bar e_i\}$,  and  $\bar 5_i
\supset \{ \bar d_i, \; L_i\}$.  For $\tan\beta \sim 1$, the top quark yukawa 
coupling is the largest yukawa coupling in the theory. It enters the 
superspace potential in the expression
\begin{eqnarray}
 W & \supset  \lambda_i^u \; ( 10_i \; 10_i \; H) &
\end{eqnarray}
where  $H$ is a 5 of SU(5) containing the Higgs doublets as well as their color triplet 
partners and we work in a basis where the up quark yukawa coupling is diagonal.
(Note, this simple SU(5) model with Higgs in the $5$ and $\bar 5$ 
representation and only dimension 4 fermion mass operators cannot fit the known
fermion masses.  Nevertheless, this is a useful exercise, since in any more 
realistic theory, the top quark yukawa coupling must still be large.\cite{bhs}
 
RG running from $M$ to $M_G$ induces lepton flavor violating masses for 
right-handed sleptons given by\cite{hkr,bhs,hmty}
\begin{eqnarray}
\delta_{ij}^{\bar e} & \sim    \;\; -  \lambda_t^2\; \delta_{3i}\;
\delta_{3j} \; \ln({M \over M_G}) &
\end{eqnarray}
In the effective theory below $M_G$, in a basis where lepton masses are now diagonal,
we have
\begin{eqnarray}
\delta_{ij}^{\bar e} & \sim    \;\;-  \lambda_t^2\; V^*_{3i}\;
V_{3j} \; \ln({M \over M_G}) &
\end{eqnarray}
where $V = V^{CKM}$.
Note, in SU(5), only right-handed sleptons are affected.  

The CKM elements
mixing the first two families with the third are small. In addition, with only 
$\delta_{ij}^{\bar e} \neq 0$, LFV is further suppressed due to a subtle  
cancellation between neutralino and higgsino contributions\cite{hmty}. As a result
lepton flavor violating processes are well within experimental bounds  and
possibly beyond the reach of future experiments.  

In SO(10), on the otherhand, both $\delta_{ij}^{\bar e}$ and 
$\delta_{ij}^{L}$ are non-zero.  Non-vanishing
contributions to $\delta_{ij}^{L}$ occur,
even in the limit of small $\tan\beta$, because both $L$ and $\bar e$
are contained in a $16 \supset \{ Q,\; \bar u,\; \bar d,\; L,\; \bar e\}$ 
of SO(10).  The combination of both these terms avoids the accidental 
cancellation discussed previously when only
$\delta_{ij}^{\bar e} \neq 0$
\cite{hmty,bhs}.  In this case, observable flavor violating effects may be expected
in future experiments.  Moreover, certain regions of parameter space are already
ruled out.   A study of the large $\tan\beta$ regime has also been carried out, see
 Ciafaloni et al.\cite{crs}, with results similar to those at low $\tan\beta$.

\subsection*{C. --- Family mass hierarchy and the Froggatt-Nielsen mechanism}

The problem with the specific GUT models discussed above is that they
give unrealistic fermion masses and mixing angles.  In order to improve upon
this situation within the context of GUTs one needs to either add several Higgs
multiplets (with some in higher dimensional representations of the GUT symmetry)
or consider the possibility of a simple Higgs sector but with higher
dimension effective fermion mass operators.  The latter case can provide effective
higher dimensional Higgs representations by incorporating direct products
such as ($5 * 24 \supset 45 + \cdots$) in SU(5).  Moreover, it was shown by Froggatt
and Nielsen\cite{fn} that these effective fermion mass operators are ``natural"
in theories with heavy intermediate states and softly broken flavor symmetries.

As an example, consider the renormalizable superspace potential given by
\begin{eqnarray}
W  & = \psi_3 \; \psi_3 \; H \;\; + \;\; \psi_2 \; \chi \; H & + \;\; \bar \chi \;
( M_{FN} \; \chi \;\; + \;\; \phi \; \psi_3 )
\end{eqnarray}
where $\psi_2 \;(\psi_3)$ represent the second (third) generation of quarks or 
leptons,  $H$ is the electroweak Higgs, ($\chi, \; \bar \chi$) are heavy
Froggatt-Nielsen states with mass $M_{FN}$ and $\phi$ contains a scalar whose
vev breaks the FN flavor symmetry at a scale below $M_{FN}$, so that $\epsilon
\equiv <\phi>/M_{FN} << 1$.  In the effective theory below $M_{FN}$, the FN
states ($\chi, \; \bar \chi$) are integrated out; giving
the effective superspace potential
\begin{eqnarray}
W  & = \psi_3 \; \psi_3 \; H \;\; + \;\; \epsilon \; \psi_2 \; \psi_3 \; H & 
\end{eqnarray}
plus calculable corrections of order $\epsilon^2$.  Thus we have a 2$\times$2 
yukawa matrix of the form
\begin{eqnarray}
\lambda & =  \left( \begin{array}{cc}   0 & \epsilon \\
                             \epsilon & 1  \end{array} \right)
\end{eqnarray}
In these theories, the scale $M_I = M_{FN}$.  

It has been shown by Dimopoulos and
Pomarol~\cite{dp} that the  FN mechanism can lead to enhanced flavor violation 
due large yukawa couplings (of order one) as well as to the mixing of heavy 
FN scalar states with light squarks and sleptons.
Consider the scalar masses
\begin{eqnarray}
{\cal L}_{soft} \supset \tilde m^2_{\psi_3} \; |\psi_3|^2 \;\; + \;\;
\tilde m^2_{\chi} \; |\chi|^2 &+ \;\; \tilde m^2_{\psi_2} \; |\psi_2|^2 \;\; + \;\;
\cdots
\end{eqnarray}
Assume that at the scale $M$ we have universal boundary conditions
\begin{eqnarray}
\tilde m^2_{\psi_3}(M) \;\;= \;\; \tilde m^2_{\psi_2}(M) & = \;\;
\tilde m^2_{\chi}(M) \;\; = \;\; m_0^2 & 
\end{eqnarray}
After RG running from $M$ to $M_{I}$ and integrating out ($\chi, \; \bar \chi$)
we obtain the scalar mass matrix for these two families given by
\begin{eqnarray}
\tilde m^2 & \sim \left( \begin{array}{cc}
                          \tilde m^2_{\psi_2}(M_{I}) & 0 \\
                            0 & \tilde m^2_{\psi_3}(M_{I}) \;\; + \;\;  \epsilon
\; \tilde m^2_{\chi}(M_{I})  \end{array} \right) & = \;
\left( \begin{array}{cc}
                          \tilde m^2_2 & 0 \\
                            0 & \tilde m^2_3  \end{array} \right)
\end{eqnarray}

Since the yukawa couplings in the renormalizable theory above $M_I$ are
assumed to be of order one, we have $(\tilde m_2^2 - \tilde m_3^2)/\tilde m_3^2 \sim
1$. When extended to three families,  order one flavor splittings between all three
families of sleptons are induced.  Such large splittings between the third and the 
first two families has already been discussed in the previous section.  It leads to
acceptable LFV rates due to the small mixing angles between the third and first two
families. Order one splittings between the first  and second family,  on the otherhand,
gives unacceptable LFV rates, since  Cabibbo like mixing between the first
two families is not small.   Recently Lucas\cite{lucas}
has calculated the LFV rates in an SO(10) SUSY GUT with realistic fermion
masses and mixing angles\cite{lr,bcrw}.  LFV interactions place severe
constraints on this model.\footnote{This analysis assumed that the messenger 
scale is the Planck scale with universal boundary conditions for soft 
SUSY breaking parameters at $M_{Pl}$.} 
Consistency with present data is only obtained with sufficiently heavy 
scalars; in particular, sneutrinos can be as light as  800 GeV, but 
only in a very restricted region of parameter space. 

\subsubsection*{U(2) family symmetry}

When using the Froggatt-Nielsen mechanism to generate a fermion
mass hierarchy, one may also need to suppress large slepton mass mixing
 between the first and second families.   This can be 
accomplished by a non-abelian family symmetry which is only broken below the 
FN scale or by lowering the messenger scale below $M_I \; (= M_{FN})$, 
such as in gauge-mediated SUSY breaking models\cite{lsgauge,dn}.\footnote{In 
gauge-mediated SUSY breaking models, the messenger scale may be as small 
as O($10^5$) GeV.  The suppression of flavor violating
effects is one of the main motivations for these models.} Several such 
symmetries have been considered in the literature.  These include: SU(2), 
SU(3), S$_3$, U(2), $\Delta(3n^2)$ with $n=4,5$. 

As an example consider the family symmetry group U(2)\cite{bdh}.  
Extensions to include an SU(5)~\cite{bh} (or SO(10) ~\cite{bhrr,bhr}) GUT 
have also been considered.  In the $SO(10)\times U(2)$ model, the first 
two families transform as a (16,2)  and the third family
transforms as a (16,1) (represented by the fields $16_a,\; a = 1,2$ and $16_3$). 

The superspace potential for the fermion mass sector is given by
\begin{eqnarray}
W & = \;\; 16_3 \; 16_3 \; 10 \;\; + \;\; 16_a \; \chi^a \; 10 & \nonumber \\
 &  \;\; + \;\; \bar \chi_a\;(M_{FN} \; \chi^a \;\; + \;\; S^{ab} \;
16_b \;\; + \;\; A^{ab} \; 16_b \;\; + \;\; \phi^a \; 16_3) &
 \end{eqnarray}
where $10$ contains the electroweak Higgs doublets and their color triplet 
partners, ($\bar \chi_a,\; \chi^a$) are the massive FN states,  and  
($S^{ab} = S^{ba},\; A^{ab} = - A^{ba}, \; \phi^a$) contain the scalars 
which spontaneously break the FN U(2) symmetry.\footnote{
Note, in order to fit fermion masses and mixing angles,  $S^{ab}$ 
transforms as a 45 while $M_{FN}$ transforms as a direct sum of 1 + 45.}  
The vacuum expectation values of the latter
fields determine the small parameters
\begin{eqnarray}
\epsilon & = & {<\phi^2> \over M_{FN}} \;  \approx \; {<S^{22}> \over M_{FN}} 
\nonumber \\
\epsilon\,' & = & {<A^{12}> \over M_{FN}}   \nonumber  \\
\epsilon\,'  & <  & \epsilon  
\end{eqnarray}

This theory results in  fermion yukawa matrices schematically given by
\begin{eqnarray}
\lambda & \sim  \;  \left( \begin{array}{ccc}  0 & \epsilon\,' & 0 \\
                                        -\epsilon\,' & \epsilon & \epsilon \\
                                       0 & \epsilon & 1  \end{array} \right) &
\end{eqnarray}
with $\epsilon \sim V_{cb} \sim 0.03$.  For more details, 
see refs.~\cite{bhrr,bhr}.

Using a simple operator analysis, the scalar mass are given by\cite{bhr}
\begin{eqnarray}
\tilde m^2 & \sim \; \left( \begin{array}{ccc}  m_1^2 & 0 & \epsilon \; 
\epsilon\,'\; m_5^2 \\
                     0 & m_1^2\; (1 \; + \; \epsilon^2) & \epsilon \; m_4^2\\
                    \epsilon \; \epsilon\,'\; m_5^2 & \epsilon \; m_4^2 & m_3^2 
\end{array} \right) &
\end{eqnarray}

Hence $\delta_{12}^{\bar e} \; \sim \; \delta_{12}^{L} \; \sim \; \epsilon^2 
\; \sim \;
10^{-3}$; consistent with experimental bounds (see eqn. ~\ref{eq:bound}).

\section*{The Bottom Line}

\underline{If the messenger scale $\bf M$ for soft SUSY breaking is above 
the ``structure}
\newline
\underline{scale" $\bf M_I$ for fermion mass hierarchies, 
then observable {\em lepton flavor violation}} \underline{is predicted due 
to the RG running of slepton masses from $\bf M$ to
$\bf M_I$.} In the examples discussed in this talk, the messenger scale 
was assumed to be the Planck scale.

$\bullet$ {\em For the ``structure scale" of fermion masses three cases were
considered:}
\indent
\begin{itemize}
\item[A.] $M_I \sim 10^{12}$ GeV  --- Neutrino masses in the MSSM, consistent 
with $m_{\nu_{\tau}} \sim $ few eV;
\item[B.] $M_I =  M_G \sim 10^{16}$ GeV --- GUTs and the third family yukawa couplings;
\item[C.] $M_I = M_{FN} \geq 10^{12}$ GeV --- Family hierarchy described by
Froggatt-Nielsen mechanism.
\end{itemize}

$\bullet$  {\em The results are strongly model dependent, but in most cases LFV
effects should be observed in the next generation experiments.}

\underline{Lepton flavor violation may be a rich goldmine of ``GUT scale" 
physics.}

\begin{center}  {\bf OR}
\end{center}

 \underline{If $M \; << \; M_I$, then lepton flavor violation may be 
 suppressed.} This would be the case in gauge-mediated SUSY breaking models 
 where $M << M_{Pl}$.

\begin{center}
{\bf ACKNOWLEDGEMENT}
\end{center}
I would like to thank K. Tobe for interesting discussions.
This work is partially supported by DOE contract DOE/ER/01545-731.

\end{document}